\documentclass[sigconf]{acmart}

\usepackage{fdiaz}

\usepackage{booktabs} 
\setcopyright{rightsretained}
\usepackage{makecell}
\usepackage{tcolorbox}
\usepackage{subcaption}
\usepackage{tablefootnote}
\usepackage{arydshln}
\usepackage{colortbl}
\usepackage{xcolor}
\usepackage{graphicx}
\usepackage{epstopdf}
\usepackage{balance}

\usepackage[utf8]{inputenc}
\usepackage{geometry}
\usepackage{tabularx}
\usepackage{multirow}
\usepackage{amsmath}

\definecolor{lightgray}{gray}{0.9}
\definecolor{Gray}{gray}{0.9}
\definecolor{highlight}{rgb}{0.9, 0.9, 0.9} 

\AtBeginDocument{%
  }

\acmYear{2026}
\setcopyright{cc}
\setcctype{by}

\acmConference[SIGIR '26] {Proceedings of the 49th International ACM SIGIR Conference on Research and Development in Information Retrieval}{July 20--24, 2026}{Melbourne, VIC, Australia.}
\acmBooktitle{Proceedings of the 49th International ACM SIGIR Conference on Research and Development in Information Retrieval (SIGIR '26), July 20--24, 2026, Melbourne, VIC, Australia}
\acmISBN{979-8-4007-2599-9/2026/07}
\acmDOI{10.1145/3805712.3808626}

\ccsdesc[500]{Information systems~Test collections}
\ccsdesc[500]{Information systems~Multilingual and cross-lingual retrieval}

\keywords{Tip-of-the-Tongue Known-Item Retrieval; Synthetic Query Generation; Multilingual Retrieval}

\begin{document}

\title{Multilingual and Domain-Agnostic Tip-of-the-Tongue Query Generation for Simulated Evaluation}

\author{Xuhong He}
\orcid{0009-0002-6360-4488}
\affiliation{%
  \institution{Carnegie Mellon University}
  \city{Pittsburgh}
  \state{PA}
  \country{USA}
}
\email{xuhongh@cs.cmu.edu}

\author{To Eun Kim}
\orcid{0000-0002-2807-1623}
\affiliation{%
  \institution{Carnegie Mellon University}
  \city{Pittsburgh}
  \state{PA}
  \country{USA}
}
\email{toeunk@cs.cmu.edu}

\author{Maik Fröbe}
\orcid{0000-0002-1003-981X}
\affiliation{%
  \institution{Friedrich-Schiller-Universität Jena}
  \city{Jena}
  \country{Germany}
}
\email{maik.froebe@uni-jena.de}

\author{Jaime Arguello}
\orcid{0000-0002-7645-0556}
\affiliation{%
  \institution{UNC Chapel Hill}
  \city{Chapel Hill}
  \state{NC}
  \country{USA}
}
\email{jarguell@email.unc.edu}

\author{Bhaskar Mitra}
\orcid{0000-0002-5270-5550}
\affiliation{%
  \institution{Independent Researcher}
  \city{Tiohtià:ke / Montréal}
  \state{QC}
  \country{Canada}
}
\email{bhaskar.mitra@acm.org}

\author{Fernando Diaz}
\orcid{0000-0003-2345-1288}
\affiliation{%
  \institution{Carnegie Mellon University}
  \city{Pittsburgh}
  \state{PA}
  \country{USA}
}
\email{diazf@acm.org}

\begin{abstract}
Tip-of-the-Tongue (ToT) retrieval benchmarks have largely focused on English, limiting their applicability to multilingual information access. In this work, we construct multilingual ToT test collections for Chinese, Japanese, Korean, and English, using an LLM–based query simulation framework. We systematically study how prompt language and source document language affect the fidelity of simulated ToT queries, validating synthetic queries through system rank correlation against real user queries. Our results show that effective ToT simulation requires language-aware design choices: non-English language sources are generally important, while English Wikipedia can be beneficial when non-English sources provide insufficient information for query generation. Based on these findings, we release four ToT test collections with 5,000 queries per language across multiple domains. This work provides the first large-scale multilingual ToT benchmark and offers practical guidance for constructing realistic ToT datasets beyond English.
\footnote{\url{https://github.com/kimdanny/ntcir-19-tot}}
\footnote{\url{https://zenodo.org/records/18777084}}
\end{abstract}
\maketitle
\section{Introduction}
\label{sec:introduction}

Tip-of-the-tongue (ToT) retrieval is a form of known-item retrieval in which searchers attempt to refind a specific item but are unable to recall a reliable identifying term \cite{arguello-movie-identification}. 
In this state, users often remember partial, indirect, or noisy information about the target item, such as contextual details, personal experiences, or incorrect attributes.
As a result, ToT queries tend to be verbose, frequently containing false memories, subjective descriptions, and exclusion criteria rather than precise identifiers.
These characteristics distinguish ToT retrieval from conventional ad hoc search and pose unique challenges for retrieval systems \cite{arguello2023overview}.

Supporting ToT retrieval is important because such queries constitute a substantial portion of real-world search behavior \cite{Meier21-complex-reddit} and are associated with high levels of user frustration when retrieval fails \cite{elsweiler2007towards}.
ToT search behavior is widely observed in community question answering (CQA) platforms such as Reddit, where users publicly articulate their refinding attempts across a wide range of topics \cite{frobe2023-performance-pred}.
This prevalence and diversity have motivated the development of benchmark datasets for ToT retrieval, most notably through shared evaluation efforts in the TREC forum \cite{arguello2023overview, arguello2024overview, arguello2025overview}.
More recently, advances in large language models (LLMs) have enabled scalable ToT query simulation, allowing evaluation to move beyond costly manual query collections \cite{he2025tot}.

Despite these efforts, existing ToT retrieval benchmarks remain limited in both language and domain coverage.
Most prior datasets and evaluation efforts are English-centric, even though ToT search behavior and retrieval failures are not confined to English-speaking users and occur broadly across CQA platforms in many languages.\footnote{For example, Naver (Korean), Yahoo! Japan (Japanese), and Baidu (Chinese) run similar CQA websites where there are constant requests on ToT item refinding.}
Furthermore, while some recent studies \cite{arguello2024overview, he2025tot} have begun to extend ToT evaluation beyond casual leisure domains, the majority of existing datasets still exhibit strong domain skewness toward entertainment-related content such as movies and music \cite{elsweiler2011casualleisure, bogers2025exploring, Bhargav23MusicTOT, arguello-movie-identification}.
These limitations restrict the generalizability of current ToT retrieval evaluations.

In light of these gaps, we propose a multilingual and domain-agnostic simulated ToT retrieval evaluation method with rigorous validation of the simulation process.
Our approach leverages LLM-based query generation to systematically expand ToT evaluation beyond English, covering three additional East Asian languages: Chinese, Japanese, and Korean (hereafter CJK).
At the same time, the proposed method is designed to span general-domain content rather than being restricted to narrow topical categories.
This combination enables broader and more realistic evaluation of ToT retrieval systems.

We empirically evaluate multiple ToT query generation strategies and validate the resulting simulations using system rank correlation analysis, following established meta-evaluation methodologies \cite{carterette09rankcorrelation, he2025tot}.
Our results show that the proposed method maintains high system rank correlation across all evaluated languages, providing evidence for the fidelity and robustness of our simulated evaluation method.
Using this method, we construct a multilingual (CJK and English), general-domain ToT retrieval test collection, which has been released for use in the upcoming NTCIR evaluation forum.

\section{Related Work}

\paragraph{\textbf{ToT Datasets}}
Prior work on tip-of-the-tongue (ToT) retrieval has produced a range of datasets spanning both single-domain and multi-domain settings. Early efforts primarily focused on well-defined content domains. In the movie domain, \citet{arguello-movie-identification} introduced a ToT dataset derived from the \textit{IRememberThisMovie} forum. Parallel efforts explored ToT phenomena in books \cite{Bhargav-2022-wsdm, lin-etal-2023-whatsthatbook}, constructing datasets from user queries on Reddit and Goodreads, and in music \cite{Bhargav23MusicTOT}, leveraging posts from the \textit{/r/tipofmytongue} subreddit. Also, ToT datasets have been extended to the video game domain, using data from the \textit{/r/tipofmyjoystick} subreddit \cite{gameTOT}.

Beyond single-domain collections, several works have pursued broader coverage across domains. \citet{Meier21-complex-reddit} introduced a multi-domain ToT dataset focused on casual leisure content extracted from Reddit. \citet{bogers:2026} compared the effectiveness of humans and large language models in answering questions accross several ToT domains.The TOMT-KIS benchmark \cite{frobe2023-performance-pred}, also derived from \textit{/r/tipofmytongue}, further emphasized known-item search under ToT conditions. Large-scale evaluation campaigns have additionally contributed multi-domain ToT resources. The TREC 2024 \cite{arguello2024overview} and TREC 2025 \cite{arguello2025overview} ToT tracks adopted a synthetic ToT query generation framework \cite{he2025tot}, enabling controlled construction of test collections across a broad set of domains. Finally, a recent effort introduced a multimodal ToT query dataset created through human annotation \cite{ch2025browsing}.

\paragraph{\textbf{TREC ToT Tracks}}

The TREC ToT tracks have played a central role in standardizing ToT evaluation. The inaugural TREC 2023 ToT track~\cite{arguello2023overview} focused exclusively on the movie domain \cite{arguello-movie-identification}. Subsequent iterations significantly broadened both methodological rigor and domain coverage. In TREC 2024 \cite{arguello2024overview} and TREC 2025 \cite{arguello2025overview}, organizers adopted rigorous principles from the known-item query simulation literature~\cite{balog2006overviewWebclef, Azzopardi07SimulatedQueries, Kim09desktop, Elsweiler2011Seeding}, emphasizing systematic synthetic query generation and validation. This line of work culminated in a modernized ToT query simulation and validation pipeline \cite{he2025tot}, grounded in system ranking correlation analyses.

Using this framework, TREC 2024 expanded beyond movies to include landmarks and celebrities, while TREC 2025 further generalized the approach, resulting in a test collection that spanning 53 domains. Despite this breadth, both tracks remained monolingual and English-centric, reflecting a broader limitation in existing ToT evaluation resources.

\paragraph{\textbf{Our Contributions}}

Our work builds on the progressive line of ToT dataset construction and evaluation established by prior TREC efforts, while introducing several differences. Similar to recent TREC tracks, we rely on a Wikipedia-based corpus and adopt a domain-agnostic synthetic ToT query generation framework. The generation process continues to follow rigorous simulation validation practices, including system ranking correlation analysis, ensuring comparability and reliability of the resulting benchmarks. 

The core contribution of this work lies in its multilingual and multi-domain design. As far as we are aware, one prior effort \cite{ch2025browsing} has explored multilingual ToT benchmarking. While valuable, that dataset is still primarily English-centric: most ToT queries are written in English, with multilinguality arising mainly from answer documents in other languages, accounting for only a minority of the data. In contrast, our dataset is multilingual by construction. Each language is associated with its own corpus, and all ToT queries are written in the target language. We cover four languages---English, Chinese, Japanese, and Korean---with 5{,}000 queries per language, spanning Movies, People, and general-domain content found in Wikipedia.

\section{Method and Experimental Setup}

To support Tip-of-the-Tongue (ToT) retrieval research in more linguistically and culturally diverse communities, we construct multilingual ToT test collections in four languages: Chinese, Japanese, Korean, and English. Each language is treated independently, with both corpus and ToT queries written in the same language, resulting in four self-contained test collections.

Our construction pipeline consists of three main stages. First, we build ToT entity sampling pools from the corresponding Wikipedia corpus (\S\ref{subsec:sampling_pool}). Second, we generate synthetic ToT queries using an LLM (\S\ref{subsec:query_gen}). Third, we validate the generated queries through system ranking correlation analysis to ensure that synthetic queries induce retrieval behavior consistent with real user ToT queries (\S\ref{subsec:query_val}). Based on this validation framework, we conduct a series of prompt variation experiments to identify effective query generation strategies for each language (\S\ref{subsec:exp}).\footnote{Since English has been the primary focus of prior ToT retrieval studies, our experiments place particular emphasis on the three East Asian languages.}

\subsection{Sampling ToT Entity Candidates}\label{subsec:sampling_pool}

We use Wikipedia as the underlying corpus due to its broad topical coverage and its rich cross-lingual linking structure. Specifically, we rely on publicly available Wikipedia dumps archived in November 2023 and distributed via HuggingFace.\footnote{\url{https://huggingface.co/datasets/wikimedia/wikipedia}} These dumps include complete English, Chinese, Japanese, and Korean  Wikipedia pages. We sample ToT entities for generation from these corpora.

Although the final retrieval corpus used for benchmarking remains identical to the original Wikipedia corpus, we partition the dataset internally to support subsequent analysis. For each East Asian language (CJK), we divide pages into two disjoint partitions:
\begin{itemize}[leftmargin=1.7em]
    \item \textit{monolingual pages}, which exist only in the target language, and
    \item \textit{bilingual pages}, which have links to the English Wikipedia page for the same entity.
\end{itemize}
This partitioning allows us to explicitly control the presence or absence of English-aligned entities during query generation. For English, we use the original, non-partitioned corpus. In total, this process yields seven distinct partitions: two partitions (monolingual and bilingual) for each of the three East Asian languages, and one English partition.

Following prior ToT dataset construction work \cite{he2025tot,arguello2024overview}, we filter each partition by popularity, retaining the top 20\% of pages ranked by page view counts, and by document length, removing entities that are overly niche or insufficiently descriptive. We then assign each page to a coarse-grained domain using Wikidata metadata, grouping entities into Movies, People, and General categories.

To ensure coverage across different familiarity levels, we perform stratified sampling by popularity within each domain. Specifically, entities are sorted by popularity, divided into 20 equal-sized buckets, and sampled uniformly to obtain 2{,}500 entities from the monolingual partition and 2{,}500 from the bilingual partition under an 8:1:1 domain ratio (80\% General, 10\% Movies, 10\% People). For English, which does not have monolingual or bilingual partitions, we directly sample 5{,}000 entities from the single unpartitioned pool using the same stratification strategy. This yields 5{,}000 ToT entity candidates per language for Chinese, Japanese, Korean, and English.

\subsection{Synthetic ToT Query Generation}\label{subsec:query_gen}
For synthetic ToT query generation, we adapt the LLM-elicitation method used by \citet{he2025tot}. For each candidate entity, we generate a synthetic ToT query using the following procedure.

\begin{enumerate}[leftmargin=1.7em]

\item \textit{Summarization of the entity's Wikipedia page}:\\
The full Wikipedia page text of the sampled entity is used to construct a summarization prompt for the GPT-4o model with temperature 0.5.\footnote{GPT4o-2024-08-06} If the page exceeds the model's token limit, it is truncated accordingly. The model produces a two-paragraph summary that captures salient attributes of the entity while abstracting away surface-level lexical details.

\item \textit{Constructing a query generation prompt}:\\
Using the generated summary as contextual input, we construct a ToT user simulation prompt that instructs the model to produce a natural language ToT query of the target entity. The same GPT-4o model is used for query generation but with temperature 0.3, following the findings of \citet{he2025tot}. We experimented with multiple prompt variants to identify the one that best approximates the system ranking derived from real ToT queries. Details of these prompt variations and their experimental evaluation are described in \S\ref{subsec:exp}.

\item \textit{Entity name anonymity check}:\\
A critical requirement for ToT query generation is that they must not explicitly reveal the target entity name. To enforce this constraint, we perform a post-generation check to detect occurrences of the entity name in the generated query. If the name is detected, we retry query generation using the same prompt, for up to three attempts. Queries that still fail this anonymity check after three retries are programmed to be discarded; however, all queries passed the check within the allowed attempts.
\end{enumerate}

\subsection{Validation of the Simulated Queries}\label{subsec:query_val}
To validate the fidelity of our simulated ToT queries, we adopt a system ranking correlation analysis framework \cite{balog2006overviewWebclef,Azzopardi07SimulatedQueries,Kim09desktop}, which has also been used in recent ToT dataset construction work \cite{he2025tot}. The core premise of this approach is that if a set of synthetic queries is a good proxy for real user queries, it should induce similar rankings of retrieval systems when evaluated using standard IR metrics (e.g., NDCG@k, MRR).

We first collect a set of real human-authored ToT queries, denoted as $Q_{\text{real}}$, by manually collecting ToT question–answer pairs from community question answering (CQA) platforms that are predominantly written in each target language: Chinese, Japanese, Korean, and English.\footnote{We do not disclose queries collected from CQA websites.} 
Our goal is to collect 150 ToT entities per language for validation purpose. Due to the relatively lower availability of ToT queries in some East Asian languages, when the collected set falls short of this target, we supplement it by translating English ToT queries into the target language using a TowerInstruct-7B-v0.2 machine translation model \cite{rei2024tower}, a winner of multiple WMT 2024 tasks.

To obtain robust and diverse system rankings, we construct a broad set of retrieval models. For each East Asian language, we construct a common pool of 22 retrievers, consisting of: (i) seven lexical retrieval models based on BM25 \cite{10.5555/188490.188561} and language modeling with Dirichlet smoothing \cite{10.1145/383952.384019}, instantiated with different hyperparameter settings; (ii) one GPT-4o-based retriever following the setup of \citet{arguello2024overview}; and (iii) fourteen multilingual dense retrievers spanning multiple architectures and checkpoint variants \cite{izacard2022unsuperviseddenseinformationretrieval,reimers-2019-sentence-bert,yang2019multilingualuniversalsentenceencoder}.  
In addition, for each language, we include five representative language-specific dense retrieval models selected based on reported performance on the corresponding language-specific MTEB leaderboard \cite{muennighoff2023mteb}. As a result, for each Chinese, Japanese, and Korean collection, we construct a total of 27 retrieval systems.

Using the real ToT queries $Q_{\text{real}}$, we evaluate all retrieval systems and rank them according to one IR metric, yielding a system ranking $R_{\text{real}}$. We then repeat the same evaluation procedure using the synthetic query set $Q_{\text{syn}}$, producing a corresponding ranking $R_{\text{syn}}$. To quantify the agreement between the two rankings, we compute both Kendall’s $\tau$ and Pearson’s $r$ correlation coefficients between $R_{\text{real}}$ and $R_{\text{syn}}$. A high correlation indicates that the synthetic queries preserve the relative ordering of retrieval systems observed under real user queries, thereby validating the quality of the simulated ToT queries.

\subsection{Experimental Setup \& Hypotheses}\label{subsec:exp}
We experiment with four prompting variations to study how an English-developed ToT query generation prompt can be adapted to multilingual settings. Since prior work has shown that a carefully designed English prompt is effective for ToT query elicitation, our experiments focus on understanding how this prompt should be adapted when generating ToT queries in other languages.

We use the following notation. $P_{en}$ denotes the original English ToT query generation prompt template developed by \citet{he2025tot}. Let $t$ be a target language. Accordingly, $q_t$ represents a ToT query written in language $t$, and $\text{Wiki}_{t}$ denotes a Wikipedia page written in language $t$. $\text{Trans}_{t}$ is a translation function that translates text into the target language $t$. $\text{inst}_{en}$ is an English-written instruction that explicitly asks the model to produce output in language $t$. The operator $\oplus$ indicates string insertion into a prompt template; for example, $P_{en} \oplus \text{Wiki}_{en}$ means that the English Wikipedia text is inserted into the appropriate slot of the English prompt.

Using this notation, we define four prompting strategies:
\begin{itemize}[leftmargin=1.7em]
    \item \textbf{Variation 1: Translated Prompt + Non-English Wikipedia}
    \[
    \text{LLM}(\text{Trans}_{t}(P_{en}) \oplus \text{Wiki}_{t}) \rightarrow q_{t}
    \]
    In this strategy, the original English prompt is translated into the target language and combined with a Wikipedia page written in the same language. The entire prompt is therefore written solely in language $t$.

    \item \textbf{Variation 2: English Prompt + Non-English Wikipedia}
    \[
    \text{LLM}((P_{en} \oplus \text{inst}_{en}) \oplus \text{Wiki}_{t}) \rightarrow q_{t}
    \]
    Here, the prompt remains in English but includes an explicit instruction to generate the output in language $t$. The Wikipedia content is provided in the target language, resulting in a bilingual prompt.

    \item \textbf{Variation 3: Translated Prompt + English Wikipedia}
    \[
    \text{LLM}(\text{Trans}_{t}(P_{en}) \oplus \text{Wiki}_{en}) \rightarrow q_{t}
    \]
    This variation mirrors Variation 1 in that the prompt is translated into the target language, but the Wikipedia content is taken from the English version of the page.

    \item \textbf{Variation 4: English Prompt + English Wikipedia}
    \[
    \text{Trans}_{t}(\text{LLM}(P_{en} \oplus \text{Wiki}_{en})) \rightarrow q_{t}
    \]
    In this approach, both the prompt and the Wikipedia content remain in English. The LLM first generates an English ToT query, which is then translated into the target language.
\end{itemize}

These prompting strategies are applied under different entity sampling scopes. Recall that for East Asian languages we construct both a \textit{monolingual} and a \textit{bilingual} partition, while the \textit{full set} refers to the unpartitioned sampling pool. Variations 1 and 2 can be applied to both the full set and the monolingual partition. In contrast, all four variations are applicable to the bilingual partition, since it provides access to both non-English language and English Wikipedia pages.

Based on the four variations, we can ask two research questions:
\begin{itemize}[leftmargin=1.7em]
    \item \textbf{RQ1}: How does the language of the query generation prompt (non-English vs.\ English) affect the fidelity of simulated ToT queries, as measured by system-rank correlation with real user queries?
    \item \textbf{RQ2}: How does the language of the source Wikipedia content (non-English vs.\ English) influence the quality of simulated ToT queries across different languages and sampling partitions?
\end{itemize}

\section{Results and Analyses}\label{sec:results}

\subsection{System Rank Correlation}

Results for the CJK languages are reported in Tables~\ref{tab:results_zh}, \ref{tab:results_ja}, and \ref{tab:results_ko}. We evaluate all sampling partitions, including the full set, which serves as a global reference for average performance across entity types. We primarily base our analysis on Kendall’s $\tau$ correlation, as linear relationship-based measures such as Pearson's $r$ are more sensitive to outliers and may yield misleading interpretations in this setting. Pearson's $r$ is reported for completeness.

For Chinese (Table~\ref{tab:results_zh}), strong rank correlations are observed across all experimental settings. In particular, English-Wikipedia-based generation in the bilingual partition achieves the highest correlations, while Chinese-language source generation with translated prompts also performs competitively on the full set.

For Japanese (Table~\ref{tab:results_ja}), overall correlation levels are lower than those observed for Chinese and Korean. Across the full and bilingual partitions, configurations using English prompts consistently outperform their Japanese-language prompt counterparts.

For Korean (Table~\ref{tab:results_ko}), correlations are uniformly high across settings. Prompt's instruction language seems to be less important, while Korean Wikipedia content achieves clear win against English Wikipedia-based generation.

To select the best query generation strategy for monolingual and bilingual partition, we compute the mean Kendall's $\tau$ across the three base retrieval metrics (NDCG@100, NDCG@1000, and MRR). The best-performing configurations according to this mean $\tau$ are boldfaced in the tables. Based on this criterion, the selected strategies are: for Chinese, Variation~1 for the monolingual partition and Variation~3 for the bilingual partition; for Japanese, Variation~2 for both monolingual and bilingual partitions; and for Korean, Variation~2 for the monolingual partition and Variation~1 for the bilingual partition.

\textbf{RQ1 (Prompt Language).}
The effect of prompt language is language-dependent rather than universal. For Japanese, English-written prompts yield higher system-rank correlations than non-English language prompts, suggesting that English instructions provide more stable or effective guidance for the LLM during ToT query generation. In contrast, for Chinese, and Korean non-English language prompts perform competitively and do not exhibit a consistent disadvantage. Overall, prompt language influences ToT simulation fidelity, but the optimal choice depends on the target language.

\textbf{RQ2 (Source Wikipedia Language).}
The impact of source Wikipedia language is also language-specific. English Wikipedia content leads to the highest correlations for Chinese in the bilingual setting, indicating that richer English articles can improve ToT simulation when non-English language resources are comparatively less detailed. However, for Japanese and Korean, non-English Wikipedia content outperforms English, confirming our observation that these corpora are sufficiently rich unlike those of Chinese, and that English content does not provide additional benefits. Consequently, the choice of source Wikipedia language should be made on a per-language basis rather than assumed to generalize across languages.

\begin{table*}[t]
  \centering
  \caption{Chinese (ZH): system rank correlation results across 8 experimental settings.}
  \label{tab:results_zh}
  \resizebox{0.8\textwidth}{!}{
  \begin{tabular}{clllcccccccc}
    \toprule
    & \multicolumn{3}{c}{\textbf{Experimental Configuration}} & \multicolumn{2}{c}{\textbf{NDCG@100}} & \multicolumn{2}{c}{\textbf{NDCG@1000}} & \multicolumn{2}{c}{\textbf{MRR}} & \multicolumn{2}{c}{\textbf{Metric Mean}} \\
    \cmidrule(lr){2-4} \cmidrule(lr){5-6} \cmidrule(lr){7-8} \cmidrule(lr){9-10} \cmidrule(lr){11-12}
    \textbf{ID} & \textbf{Candidate Partition} & \textbf{Prompt} & \textbf{Wiki} & \textbf{$\tau$} & \textbf{$r$} & \textbf{$\tau$} & \textbf{$r$} & \textbf{$\tau$} & \textbf{$r$} & \textbf{$\bar{\tau}$} & \textbf{$\bar{r}$} \\
    \midrule
    1 & Full Set & Chinese (Trans) & Chinese & 0.7805 & 0.8647 & 0.7880 & 0.8823 & 0.6390 & 0.7852 & 0.7358 & 0.8441 \\
    2 & Full Set & English & Chinese & 0.6562 & 0.6967 & 0.6963 & 0.7395 & 0.5989 & 0.6135 & 0.6505 & 0.6832 \\
    \midrule
    3 & Monolingual & Chinese (Trans) & Chinese & 0.7040 & 0.8159 & 0.6963 & 0.8784 & 0.6370 & 0.7124 & \textbf{0.6791} & 0.8022 \\
    4 & Monolingual & English & Chinese & 0.6313 & 0.6483 & 0.6848 & 0.7521 & 0.5415 & 0.5288 & 0.6192 & 0.6431 \\
    \midrule
    5 & Bilingual & Chinese (Trans) & Chinese & 0.6619 & 0.8717 & 0.7098 & 0.8772 & 0.6218 & 0.8167 & 0.6645 & 0.8552 \\
    6 & Bilingual & English & Chinese & 0.5931 & 0.6987 & 0.6332 & 0.7266 & 0.6178 & 0.6313 & 0.6147 & 0.6855 \\
    7 & Bilingual & Chinese (Trans) & English & 0.7650 & 0.9222 & 0.7364 & 0.9306 & 0.7421 & 0.8971 & \textbf{0.7478} & 0.9166 \\
    8 & Bilingual & English & English & 0.7536 & 0.9164 & 0.7479 & 0.9210 & 0.6676 & 0.9150 & 0.7230 & 0.9175 \\
    \bottomrule
  \end{tabular}
  }
\end{table*}

\begin{table*}[t]
  \centering
  \caption{Japanese (JA): system rank correlation results across 8 experimental settings.}
  \label{tab:results_ja}
  \resizebox{0.8\textwidth}{!}{
  \begin{tabular}{clllcccccccc}
    \toprule
    & \multicolumn{3}{c}{\textbf{Experimental Configuration}} & \multicolumn{2}{c}{\textbf{NDCG@100}} & \multicolumn{2}{c}{\textbf{NDCG@1000}} & \multicolumn{2}{c}{\textbf{MRR}} & \multicolumn{2}{c}{\textbf{Metric Mean}} \\
    \cmidrule(lr){2-4} \cmidrule(lr){5-6} \cmidrule(lr){7-8} \cmidrule(lr){9-10} \cmidrule(lr){11-12}
    \textbf{ID} & \textbf{Candidate Partition} & \textbf{Prompt} & \textbf{Wiki} & \textbf{$\tau$} & \textbf{$r$} & \textbf{$\tau$} & \textbf{$r$} & \textbf{$\tau$} & \textbf{$r$} & \textbf{$\bar{\tau}$} & \textbf{$\bar{r}$} \\
    \midrule
    1 & Full Set & Japanese (Trans) & Japanese & 0.4631 & 0.9437 & 0.5183 & 0.9329 & 0.3748 & 0.9613 & 0.4521 & 0.9460 \\
    2 & Full Set & English & Japanese & 0.5611 & 0.8906 & 0.5940 & 0.8940 & 0.4833 & 0.8993 & 0.5461 & 0.8946 \\
    \midrule
    3 & Monolingual & Japanese (Trans) & Japanese & 0.4680 & 0.8659 & 0.5426 & 0.8524 & 0.4331 & 0.8998 & 0.4812 & 0.8727 \\
    4 & Monolingual & English & Japanese & 0.4986 & 0.7758 & 0.5871 & 0.8194 & 0.4733 & 0.7560 & \textbf{0.5197} & 0.7837 \\
    \midrule
    5 & Bilingual & Japanese (Trans) & Japanese & 0.4332 & 0.9205 & 0.5222 & 0.9082 & 0.3602 & 0.9349 & 0.4385 & 0.9212 \\
    6 & Bilingual & English & Japanese & 0.5358 & 0.9118 & 0.5989 & 0.9014 & 0.4914 & 0.9263 & \textbf{0.5420} & 0.9132 \\
    7 & Bilingual & Japanese (Trans) & English & 0.4136 & 0.9112 & 0.5316 & 0.9031 & 0.3902 & 0.9320 & 0.4451 & 0.9154 \\
    8 & Bilingual & English & English & 0.4878 & 0.9366 & 0.5645 & 0.9384 & 0.5086 & 0.9542 & 0.5203 & 0.9431 \\
    \bottomrule
  \end{tabular}
  }
\end{table*}

\begin{table*}[t]
  \centering
  \caption{Korean (KO): system-rank correlation results across 8 experimental settings.}
  \label{tab:results_ko}
  \resizebox{0.8\textwidth}{!}{
  \begin{tabular}{clllcccccccc}
    \toprule
    & \multicolumn{3}{c}{\textbf{Experimental Configuration}} & \multicolumn{2}{c}{\textbf{NDCG@100}} & \multicolumn{2}{c}{\textbf{NDCG@1000}} & \multicolumn{2}{c}{\textbf{MRR}} & \multicolumn{2}{c}{\textbf{Metric Mean}} \\
    \cmidrule(lr){2-4} \cmidrule(lr){5-6} \cmidrule(lr){7-8} \cmidrule(lr){9-10} \cmidrule(lr){11-12}
    \textbf{ID} & \textbf{Candidate Partition} & \textbf{Prompt} & \textbf{Wiki} & \textbf{$\tau$} & \textbf{$r$} & \textbf{$\tau$} & \textbf{$r$} & \textbf{$\tau$} & \textbf{$r$} & \textbf{$\bar{\tau}$} & \textbf{$\bar{r}$} \\
    \midrule
    1 & Full Set & Korean (Trans) & Korean & 0.7626 & 0.9668 & 0.8127 & 0.9696 & 0.6589 & 0.9746 & 0.7447 & 0.9703 \\
    2 & Full Set & English & Korean & 0.7133 & 0.9505 & 0.7683 & 0.9516 & 0.6377 & 0.9554 & 0.7064 & 0.9525 \\
    \midrule
    3 & Monolingual & Korean (Trans) & Korean & 0.5703 & 0.8413 & 0.6119 & 0.8690 & 0.5471 & 0.7689 & 0.5764 & 0.8264 \\
    4 & Monolingual & English & Korean & 0.6045 & 0.8240 & 0.7054 & 0.8754 & 0.5486 & 0.7036 & \textbf{0.6195} & 0.8010 \\
    \midrule
    5 & Bilingual & Korean (Trans) & Korean & 0.7989 & 0.9600 & 0.8184 & 0.9672 & 0.5690 & 0.9649 & \textbf{0.7288} & 0.9640 \\
    6 & Bilingual & English & Korean & 0.7191 & 0.9467 & 0.7931 & 0.9486 & 0.5876 & 0.9521 & 0.6999 & 0.9491 \\
    7 & Bilingual & Korean (Trans) & English & 0.7422 & 0.9496 & 0.7723 & 0.9516 & 0.6143 & 0.9577 & 0.7096 & 0.9530 \\
    8 & Bilingual & English & English & 0.7636 & 0.9370 & 0.7723 & 0.9432 & 0.6434 & 0.9551 & 0.7264 & 0.9451 \\
    \bottomrule
  \end{tabular}
  }
\end{table*}

\vspace{-10pt}
\subsection{Domain-Level Analysis}
To examine how simulation fidelity varies across entity types, we analyze validation performance by domain. Across all three languages, the General domain consistently achieves high system-rank correlations, with Kendall’s $\tau$ frequently exceeding $0.7$. For Chinese, the Movies domain yields the highest correlations among all domains. In contrast, for Japanese and Korean, both the Movies and People domains exhibit slightly lower correlations than the General domain.

\begin{table}[ht]
\centering
\setlength{\tabcolsep}{4pt} 
\caption{Statistics of each CJK ToT Test Collection.}
\begin{tabular}{l|l}
\toprule
\textbf{Category} & \textbf{Distribution (Count / Percentage)} \\
\midrule
\textbf{Total Queries} & 5,000 per language (15,000 total) \\
\textbf{Linguistic Split} & Monolingual (2,500), Bilingual (2,500) \\
\textbf{Domain Split} & Movies (10\%), People (10\%), General (80\%) \\
\textbf{Dataset Split} & Train (80\%), Dev (10\%), Test (10\%) \\
\textbf{Corpus Size} & C (1.38M), J (1.39M), K (648K) \\
\bottomrule
\end{tabular}
\label{tab:datasets_stats}
\end{table}

\subsection{Constructing Test Collections}
Using the selected query generation strategies, we generate ToT queries for the 5{,}000 entity candidates per language. For each CJK language, if an entity belongs to the monolingual partition, we apply the strategy corresponding to the highest mean $\tau$ among Exp~3 or Exp~4. If an entity belongs to the bilingual partition, we apply the best-performing strategy among Exp~5, Exp~6, Exp~7, or Exp~8, as determined by mean $\tau$.
For English, we directly follow the original English ToT query generation procedure proposed by \citet{he2025tot}, generating queries from 5{,}000 sampled entities without partitioning.

Dataset statistics for each CJK collection are summarized in Table~\ref{tab:datasets_stats}. Together, these procedures result in four final ToT test collections: Chinese, Japanese, Korean, and English.

\section{Discussion}\label{sec:discussion}

\paragraph{\textbf{Resource Contribution}}
This work makes a concrete contribution to the community by releasing a new multilingual Tip-of-the-Tongue (ToT) dataset and the accompanying infrastructure needed to support future research. Specifically, we release the training and development sets for all supported languages together with this paper, along with the full source code used to generate the dataset and supporting resources.
To align with the established evaluation practices, we plan to release the test sets following the official NTCIR-19 timeline. This schedule places the test set release close to the SIGIR 2026 conference dates, enabling coordinated evaluation, benchmarking, and community-wide participation.

\paragraph{\textbf{Availability}}
The released resources are fully available to reviewers and researchers at the time of submission. The training and development sets for all languages, along with their associated corpora, are publicly released and accessible via a permanent archival link.\footnote{\url{https://zenodo.org/records/18777084}} In addition, comprehensive documentation describing the dataset construction, statistics, and usage is provided through the official track webpage.\footnote{\url{https://ntcir-tot.github.io}} The test sets for all languages will be released during the summer in accordance with the NTCIR-19 evaluation schedule. To support transparency and reproducibility, the source code used for dataset generation and processing is also publicly available through an open-access repository.\footnote{\url{https://github.com/kimdanny/ntcir-19-tot}} All released resources are distributed under open licensing terms, allowing unrestricted use by both academic researchers and industry practitioners.

\paragraph{\textbf{Utility}}
The dataset is designed to be straightforward to use. Detailed documentation of the dataset provenance, preprocessing, and aggregation steps are provided in this paper and on the track webpage. As a result, users with standard information retrieval or natural language processing expertise should be able to adopt the resource without specialized domain knowledge. To further lower the barrier to entry, we provide usage examples through the README file in the source code repository. These examples demonstrate how to load the dataset and reproduce baseline experiments. In addition, the dataset is integrated into the well-established \texttt{ir\_datasets} Python framework, enabling practitioners to load the data with a single line of Python code and seamlessly incorporate it into existing IR pipelines.

\paragraph{\textbf{Novelty and Predicted Impact}}

This resource extends a well-established line of work on Tip-of-the-Tongue queries \cite{arguello-movie-identification} and known-item retrieval \cite{balog2006overviewWebclef, Azzopardi07SimulatedQueries, Kim09desktop} by broadening its scope to multilingual and general-domain synthetic ToT query generation for simulated evaluation and benchmarking. As such, the contribution opens new opportunities for studying ToT retrieval under multilingual, cross-lingual, and culturally diverse conditions.

ToT queries themselves fall under the broader category of complex information needs, which still remain challenging for modern retrieval systems \cite{killingback2025benchmarking}.  The proposed dataset therefore provides a valuable testbed for advancing research on complex query understanding and retrieval. 

Moreover, multilingual retrieval \cite{enevoldsen2025mmteb} remains an active and growing research area, as also evidenced by recent competition \cite{wsdmcup2026}. Our dataset allows researchers to easily construct multilingual and cross-lingual retrieval benchmarks by merging corpora across languages. Because the corpus is partitioned not only by language but also by the existence of corresponding English pages, researchers can naturally define relevance across languages, including settings where multiple bilingual documents are considered relevant for a single query.

The dataset also supports research on multicultural and multilingual LLM development. Multilinguality inherently encompasses cultural variation \cite{aronin2004exploring}, and Tip-of-the-Tongue expressions may reflect culturally specific descriptions and associations. The same underlying concept may be described in systematically different ways across Chinese-, Japanese-, Korean-, and English-speaking populations \cite{bhatt-diaz-2024-extrinsic}, making ToT queries a particularly rich signal for studying cultural variation in language use. 

Taken together, these factors suggest that the dataset has long-term value and that the community of researchers using it is likely to grow over time.

\paragraph{\textbf{Implications of Results}}
Our results demonstrate that multilingual ToT query simulation requires language-specific design choices rather than uniform reuse of English-centric pipelines. The analyses for RQ1 and RQ2 show that both prompt language and source Wikipedia language substantially influence simulation fidelity, but their effects vary across languages and entity types. Non-English language Wikipedia content is generally essential for accurately capturing language- and culture-specific ToT behavior, while English Wikipedia can be beneficial when non-English resources are less detailed. These findings suggest that effective multilingual ToT dataset construction should rely on adaptive, language-aware prompting and content selection strategies instead of direct translation or one-size-fits-all approaches.

\paragraph{\textbf{Limitation and Future Work}}
One limitation of the current dataset is its focus on East Asian languages. However, the underlying methodology for query generation is not language-specific and could be extended to other linguistic and cultural contexts. Expanding the dataset to additional languages therefore represents a natural direction for future work. In addition, modern information access systems are increasingly interactive and multi-turn in nature \cite{swirl_18}. Tip-of-the-Tongue experiences in real-world settings often unfold over multiple interactions rather than a single query \cite{Meier21-complex-reddit}. Developing multi-turn ToT retrieval system and test collections is an important and promising direction, and we view the current resource as a foundational step toward supporting such future evaluations.
\section{Conclusion}\label{sec:conclusion}
We present the first large-scale multilingual Tip-of-the-Tongue retrieval benchmark for Chinese, Japanese, and Korean, alongside a comparable English collection. Through systematic validation, we show that realistic ToT simulation requires language-aware design choices in prompt and source selection. These datasets will be used as official test collections in NTCIR-19, supporting future research on multilingual and cross-lingual ToT retrieval.

\begin{acks}
We thank Kimihiro Hasegawa and Masao Someki for Japanese CQA query collection.
\end{acks}
    
\vfill\eject
\bibliographystyle{ACM-Reference-Format}
\balance
\bibliography{XX-references}


\end{document}